\documentstyle[twocolumn,aps,epsf]{revtex}

\newcommand{\ignore}[1]{}

\begin{document}
\twocolumn[
\hsize\textwidth\columnwidth\hsize\csname @twocolumnfalse\endcsname
\draft
\title{A model for single electron decays from a strongly isolated 
 quantum dot }
\author{J. Martorell}
\address {Dept.
  d'Estructura i Constituents de la Materia, Facultat F\'{\i}sica,\\
   University of Barcelona, Barcelona 08028, Spain
}
\author{D. W. L. Sprung and P. A. Machado}
\address{
  Department of Physics and Astronomy, McMaster University\\
  Hamilton, Ontario L8S 4M1 Canada
}
\author{C.G. Smith}
\address{
  Semiconductor Physics Group, Cavendish Laboratory, Madingley Road\\
  Cambridge CB3 OHE, U K 
}
\date{\today}
\maketitle

\begin{abstract}
Recent measurements of electron escape from a non-equilibrium 
charged quantum dot are interpreted within a 2D separable model. The 
confining potential is derived from 3D self-consistent 
Poisson-Thomas-Fermi calculations. It is found that the sequence of 
decay lifetimes provides a sensitive test of the confining potential 
and its dependence on electron occupation. 
 \end{abstract}
\narrowtext
]

\section{Introduction}
\begin{figure} 
\leavevmode
\epsfxsize=8cm
\epsffile{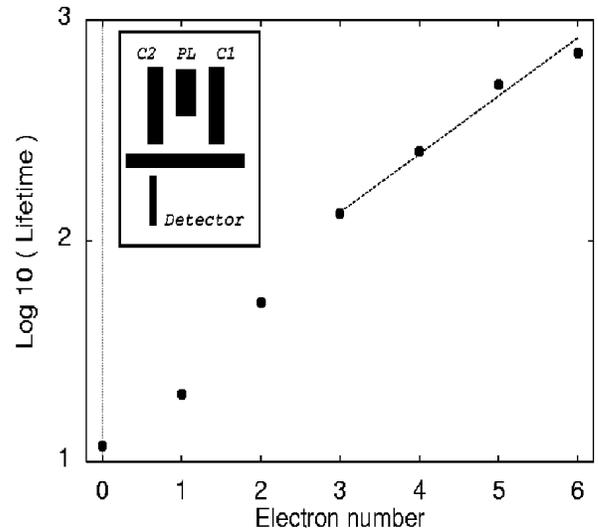}
\caption{ Experimental lifetimes (in seconds) extracted from the 
decay sequences, as reported ref. [1]. The insert shows the gate 
arrangement which defines the dot.} 
 \label{fig1} 
\end{figure}

In a recent experiment \cite{CSR99}, a strongly isolated quantum dot 
was charged with  excess electrons, and their sequential escapes were 
recorded over a one hour time period. This was repeated 150 times to 
obtain a statistical distribution of decay times. 
The dot is formed in an electron gas located at a depth of  $70 \ nm$ 
in a $GaAs-AlGaAs$ heterostructure. Its shape is defined by 
electrostatic confinement using a set of gates, as sketched in the insert 
to Fig. \ref{fig1} . The gate voltages were ramped up quickly, so 
that the dot retained a sizeable number of excess electrons when it 
was well isolated from the surrounding electron gas. The observations 
correspond to sequential tunnelling of (seven) electrons from the dot 
to the surroundings. The lifetimes extracted from the escape times 
distribution \cite{CSR99} are shown in Fig. \ref{fig1}. A striking
quasi-linear dependence of the logarithm of the lifetime on electron 
number is apparent. 

Sequential decays have been known and studied for over a century in 
the context of nuclear physics. The combined instances of alpha and 
beta decays from the heaviest elements are  responsible for most 
natural radioactivity. The description of alpha decay in terms of 
tunnelling of alpha particles through a confining potential dates 
back to the 1920's (Gamow \cite{G28}, Condon and Gurney \cite{CG28}). 
Although the basic nature of the decay as a barrier penetration is 
well understood, accurate predictions for radioactive lifetimes are 
difficult because the process by which the escaping alpha-particle is 
preformed within the nucleus requires an understanding of four-body 
correlations.  As a result, it is impossible to deduce accurate 
information on the barrier shape. Nevertheless, alpha decays have 
provided useful information on nuclear radii and the range and gross 
features of the nuclear interaction \cite{PB75}. 

It has become commonplace to say that a quantum dot is an artificial 
atom, but in fact the self-consistent potential confining electrons 
in a large dot has more in common with the mean field potential in a 
heavy nucleus: flat in the interior, with abrupt walls. An artificial 
nucleus is a more apt description, as will become clear in this 
paper. Indeed, the detection of sequential decays from an isolated 
quantum dot is a more favourable situation for study of the decay 
process, as the question of preforming the electron does not arise. 
Hence, we can more confidently test our knowledge of the confining 
barriers for electrons, as well as the profile, and 
dependence on occupation number, of the dot potential. We 
will analyze these aspects in this work, and show that these 
measurements of the lifetimes of ``radioactive quantum dots'' 
introduce new constraints on our ability to model their structure. 

The present experiment has another significant advantage 
over nuclear decays: instead of counting incoherent decays 
from a large sample of identical nuclei, here a single dot is 
involved, and the correlation between consecutive events can be 
analyzed.  In addition, it should be possible to design the shape, 
density and excitation energy of the dot within rather broad margins, 
so that future experiments on mesoscopic systems will be much more 
flexible than those in nuclear systems, where only those nuclei 
existing in nature, or created in sufficient numbers, can be 
studied. Thus, the study of electron decays from a quantum dot has 
the potential to reveal new features of the tunnelling process. This 
is a topic of currently renewed interest: see for example van Dijk 
and Nogami \cite{vDN99}. The type of simple model developed in this 
paper can be of great utility in such future studies. 

In this work we will describe the decay process using analytic 
models which incorporate characteristics of the confinement potential 
extracted from realistic numerical simulations. As the dot contains 
about 300 electrons, Poisson-Thomas-Fermi calculations should be 
adequate to describe the electron density and the confining potential 
of the dot. With these in hand we have developed accurate analytic 
approximations for the confining potential that allow us to construct 
an envelope approximation wavefunction for the electrons in the dot, 
and to compute the electron lifetimes from a fully quantal expression 
for the transmission amplitude across the barrier. 

Previous works which model a quantum dot have been concerned with the 
wave functions of confined states in the dot, the electron density 
distribution and the shape of the confining potential. For such 
purposes, only the inside of the barrier matters. It is when one 
looks at the escape of electrons from the dot that the barrier 
height, its width and shape become important; these are the new 
features explored in this paper. In section II we describe the 
development of our model, while in Section III we discuss the results 
for the sequence of lifetimes and compare them with experiment. Some 
details are relegated to two appendices. 

\section{ Modelling of Isolated Dot Decays}
\subsection{Framework}

The Poisson-Thomas-Fermi modelling is described in more detail
elsewhere \cite{MS98}, so here we list only the main steps:\\
{\it i)} first,  Poisson-Schr\"odinger (PS) and Poisson-Thomas-Fermi
(PTF) simulations as described in \cite{MS94} are performed for the
ungated heterostructure. Our inputs for the PS simulation are the
thickness and composition of each layer in the heterostructure,
and the dopant concentration in the donor layer. From these we
predict the density of the 2DEG. The only adjustable parameter is the
donor ionization energy which is set to be $e\Phi_i = 0.12 \ eV$,
in order to reproduce the measured 2DEG density, $ n_e = 2.74 \, 
10^{11} \ cm^{-2} $. For the simpler Poisson-Thomas-Fermi scheme we
employ a common relative permitivity $\varepsilon_r = 12.2$ for all
layers of the heterostructure, which, combined with the
parameters already used for the PS simulation, also reproduces the
experimental $n_e$. After  this ``fitting'' the model has no  other
free parameters.\\ 
{\it ii)} For the gated structure we use the gate layout and voltages 
of the experiment. To solve the Poisson equation for the gated 
heterostructure one has to impose as a boundary condition the value 
of the electrostatic potential on the exposed surface of the 
heterostructure, and on the gates. We assume Fermi level pinning and 
choose the energy of the surface states as the zero of the energy scale. 
In this convention, the conduction band edge is set at $e V_s = 0.67 
\ eV $ on the exposed surface. Under each gate the conduction band is 
set at  $ eV_{ms}+ eV_g$, where $V_g$ is the gate voltage and the 
metal semiconductor contact potential, $eV_{ms}$, is taken as $0.81 \ 
eV$ \cite{Mo95}. The electrostatic potential due to the gates is then 
computed using semi-analytic expressions based on the work of Davies 
{\it et al.} \cite{Davies88} and \cite{DLS95}. Added to this are: 
{\it a)} the Coulomb potential (direct term) between the electrons,
and a mirror term which imposes the boundary conditions at the
surface, and {\it b)} the contribution from the fully ionized donor
layer and its mirror term (see Sect. IIA of \cite{MWS94} for details
of a similar example.)  We neglect exchange and correlation effects, 
which are small.
\\
{\it iii)} The connection between the confining potential defined by
the conduction band edge and the electron density is completed by
using the Thomas-Fermi approximation at zero temperature:
 \begin{equation}
\rho_e({\vec r}) = {1\over {3\pi^2}} \left(  {{2m^*}\over \hbar^2}
               (E_F -eV({\vec r}))\right)^{3/2}
\label{eq:1}
\end{equation}
The PTF iteration is performed starting from the ungated
heterostructure densities as trial values.

\subsection{Equilibrium dot}

\begin{figure}
\leavevmode
\epsfxsize=8cm
\epsffile{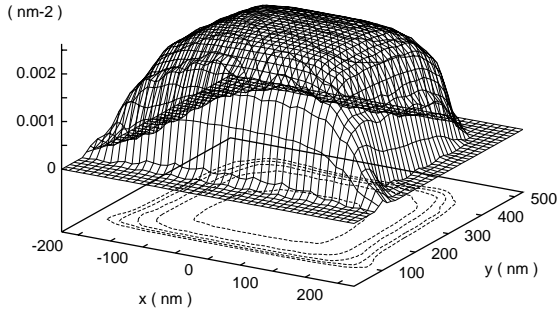}
\caption{ The two dimensional PTF density, $n_e(x,y)$, for a dot in 
equilibrium with the surrounding 2 DEG.} 
 \label{fig2}
\end{figure}

As a first step, we examine the dot in its final state after
all the excess electrons have escaped. This  corresponds to a PTF
simulation with the  same Fermi level, $E_{F,dot} = 0$, for the electrons
in the  dot and in the 2DEG outside the barriers.
The gate voltages are  taken from ref. \cite{CSR99} as $V_{PL} = -0.40
\ V$, $V_{C1}= V_{C2} = -0.44 \ V$ and $V_H = -0.7 \ V$.
The predicted PTF 3D electron distribution $\rho_e(x,y,z)$ is more
conveniently visualized in terms of a projected 2D density:
 \begin{equation}
n_e(x,y) = \int_{z_j}^{\infty} \rho_e(x,y,z) \ dz \ ,
\label{eq:2}
\end{equation}
where $z_j$ is the junction plane. The $n_e(x,y)$ 
distribution, shown  in Fig. \ref{fig2},  has an approximately 
rectangular boundary, and its maximum value is close to the 2DEG 
density of the ungated heterostructure. In this calculation the dot 
contains 286 electrons. 

\subsection{Dot with excess electrons} 

To study these configurations we set the Fermi level inside the dot, 
$E_{F,dot}$, higher than its value outside the barriers, 
$E_{F,2DEG}=0$. We can do so because the dot is well pinched off from the 
surrounding electron gas. We ran PTF simulations with equally spaced 
values for $E_{F,dot}$ running from $0$ to $17.5 \ meV$ in steps of 
$2.5 \ meV$. The occupation $Q$ of the dot increases 
linearly with $E_{F,dot}$ at the rate $2.75$ electrons per meV, 
giving occupations $286 \le Q \le 334$. 

The simulations also produce the confining potential for the 
electrons in the dot, $e V(x,y,z)$. To reduce this to a 
two-dimensional function, $U(x,y)$, we take a  weighted average over 
the density profile in the $z$ direction: 
 \begin{equation}
U_{PTF}(x,y) = {{\int_{z_j}^{\infty} e V(x,y,z) \ P(z) \ dz}\over
{\int_{z_j}^{\infty} P(z) \ dz}}
\label{eq:3}
\end{equation}
where 
 \begin{equation}
P(z) = \int_\Omega \rho_e(x,y,z)  \ dx \ dy  \ .
 \label{eq:4}
\end{equation}
Here the domain of integration $\Omega$ is
a rectangle in the $x y$ plane which extends a short distance into
the surrounding electron gas, (from $(x_l,y_l) = (-510 \ nm, -255 \
nm)$ to $(x_r,y_r)  = (510 \ nm, 255 \ nm)$.) This includes  an area
outside the dot where the 2DEG is still depleted by 
the gates. Although the computed  $V(x,y,z)$ is not separable,
previous experience with Poisson-Schr\"odinger simulations of wires
\cite{MWS94},\cite{MS96}  and circular dots, has shown us that the 
factorization ansatz leads to very good approximations when the $z$
degree of freedom is integrated out as in eq. \ref{eq:3}. 
This prescription to construct the 2D potential 
avoids the type of {\it ad hoc} assumptions often made.

\begin{figure}
\leavevmode
\epsfxsize=8cm
\epsffile{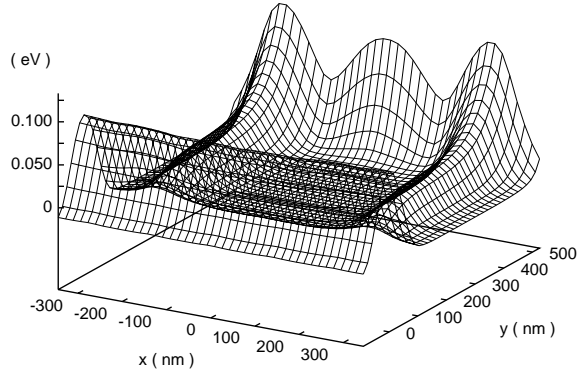}
\caption{ Two dimensional confining potential, $U_{PTF}(x,y)$
for the dot of Fig. 1.}
\label{fig3}
\end{figure}

In Fig. \ref{fig3} we show the $U_{PTF}(x,y) $ corresponding to the
equilibrium dot of Fig. \ref{fig2}. As expected from the gate layout 
shown in the inset to Fig. \ref{fig1}, it has two very high
barriers running parallel to the $x$ axis, one centered at $ y=0 $
and the other that begins with a steep rise at $y \simeq 400 \ nm$ 
(and shows clearly the mark of the three-fingered gate layout 
labelled $C1,\, C2$, and $PL$unger in Fig. \ref{fig1}). 
Tunnelling across these barriers is negligible. In addition there is a 
symmetric pair of barriers running parallel to the $y$ axis, with 
maxima at $x  \simeq \pm 238 \ nm  $ through which the electrons {\it 
do} tunnel. In the interior, the potential is 
practically constant. Although these $x$-barriers have somewhat 
increasing height with increasing $y$, the rectangular shape of the 
potential suggests using a separable approximation in cartesian 
coordinates: 
 \begin{equation}
 U_{PTF} (x,y) \simeq U_s(x,y) = U(x) +  W(y)
\label{eq:5}
\end{equation}

\begin{figure}
\leavevmode
\epsfxsize=8cm
\epsffile{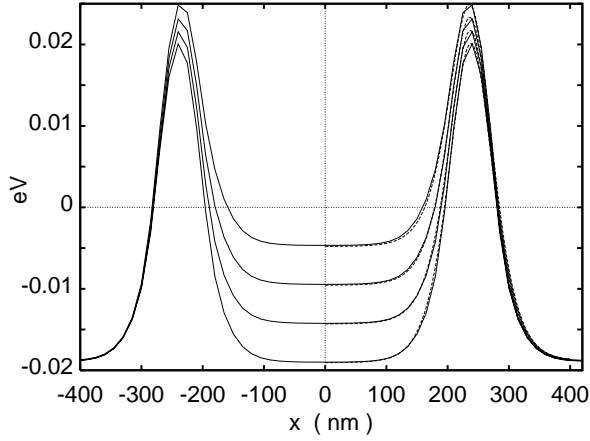}
\caption{ Continuous lines: sections at $y = 200 \ nm$ of the
$U_{PTF}(x,y)$ corresponding to $E_{F,dot} = 0.0, 0.005, 0.010$ and $
0.015 \ eV$. Dashed lines: analytic parametrization for $U(x)$ as
described in the text (the latter shown only for $x>0$ for clarity.)}
\label{fig4}
\end{figure}

We will interpret the experimental decay data using this separability 
ansatz. For the $W(y)$ barriers, which  are basically impenetrable, 
we use two simple models described below.  As a guide to a realistic 
choice for the $x$-dependent term we examine in Fig. \ref{fig4} the 
profiles of $U_{PTF}(x,y)$ at a fixed value of $y = 200 \ nm$ in the 
middle of the dot.  The profiles shown cover a range of occupations 
of up to forty excess electrons. In this range, the potential at the 
dot center increases linearly with $Q$, according to 
 \begin{equation}
U_0 = 0.347 Q - 118.4 \ meV \ . 
\label{eq:6}
\end{equation}
At large distances outside the dot, $U_\infty = -18.8$ meV is 
constant. Similarly, the location of the barrier maximum and its 
height can be parametrized as: 
 \begin{eqnarray}
x_b &=& 238 - {{ Q - 286 }\over 16} \ nm \nonumber \\
U_b &=& 0.117 Q - 13.4 \ meV \ .
\label{eq:7}
\end{eqnarray}
Note that $ d U_b / d Q \approx 1/3 d U_0 / dQ$ reflects 
the decrease of the screened Coulomb repulsion away from
the center of the dot. Furthermore, we have found that the
$x$-dependence can be very well reproduced (see Fig. \ref{fig4}) 
using the following analytic model:
 \begin{eqnarray}
U(x) &=& U_b + U_{MF}(x) \, , \quad x > 0, \nonumber \\
     &=& U(-x) \, , \qquad \qquad x < 0, \quad {\rm where} \nonumber \\
U_{MF} &\equiv& U_c {{\sinh^2 \left({{x-x_b}\over w_b}\right)}\over
{\cosh^2\left({{x-x_b}\over w_b} - \mu\right)}}
\label{eq:8}
\end{eqnarray}
This potential form has the great advantage that the transmission
coefficient for $U_{MF}$ is known analytically \cite{MF53}. $U_{MF}$
is an asymmetric barrier which takes one value for $x << x_b$ and
another value for $x >> x_b$:
 \begin{eqnarray}
U_{MF}(x_b) &=& 0 \nonumber \\
U_{MF}(\infty) &\equiv& \lim_{x \to \infty} U_{MF}(x) = U_c e^{2\mu}
\nonumber \\
U_{MF}(-\infty) &\equiv& \lim_{x \to -\infty} U_{MF}(x) =  U_c e^{-2\mu}  \ .
\label{eq:9}
\end{eqnarray}
The parameters $U_b, U_c, \mu, x_b, w_b$ allow one to fit the
barrier height, the potential floors inside and outside the dot, the
barrier spacing and the barrier width. Since the barriers are spread
quite far apart, in practice $x_b >> w_b$, so $U_{MF}(x=0) \approx
U_{MF}(-\infty)$. In this case, 
 \begin{eqnarray}
U_{0} &\equiv& U(0) \approx U_b + U_c e^{-2\mu} \nonumber \\
U_{\infty} &\equiv& \lim_{x \to \infty} U(x) = U_b + U_c e^{2\mu} .
\label{eq:10}
\end{eqnarray}
Then we can solve for
 \begin{eqnarray}
\mu &=& {1\over 4} \ln \left({{U_b - U_{\infty}}\over {U_b - U_0}}\right)
\qquad {\rm and } \nonumber \\
U_c &=& -(U_b -U_0) e^{2\mu} \, .
 \label{eq:11}
\end{eqnarray}
To determine the parameters appearing in eq. \ref{eq:8}, we
take the values of the PTF potential at the origin, $U_0$, well beyond
the barrier, $U_{\infty}$, and the value $U_{x_b}$ at the barrier
maximum $x=x_b$,  and then plot $U(x)$ to find the best $w_b$, which
turned out to be $48 \ nm$.  This gives a convenient analytic 
form for the confining potential, motivated by PTF, whose 
transmission coefficient is: 
 \begin{equation}
T = {{2 \sinh (\pi k_+) \sinh(\pi k_-)}\over {\cosh(\pi(k_+ + k_-))
          + \cosh(\pi \beta)}} \quad ,
\label{eq:12}
\end{equation}
where
\begin{eqnarray}
k_{-/+} &=& \sqrt{ {{2m^*}\over \hbar^2}(E- U_{0/\infty}) w_b^2} 
\qquad {\rm and} \nonumber \\
\beta &=& \sqrt{{{2m^*}\over \hbar^2}(2U_b-2U_c-U_0-U_{\infty})w_b^2-1}
\quad .
\label{eq:13}
\end{eqnarray}

\begin{figure}
\leavevmode
\epsfxsize=8cm
\epsffile{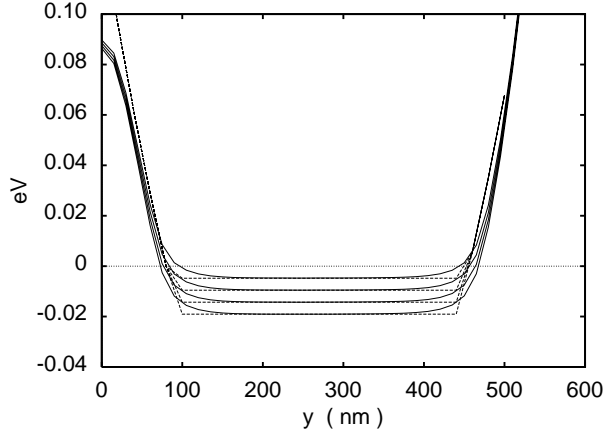}
\caption{Continuous lines: sections at $x = 0 \ nm$ of the
$U_{PTF}(x,y)$ for the same $E_{F,dot}$ as in Fig. 3. Dashed lines:
analytic parametrization of $W_{tho}(y) + U_0$ as described in the
text.}
\label{fig5}
\end{figure}

{\it Barrier shape $W(y)$}:
In Fig. \ref{fig5} we examine a section of $U_{PTF}(x=0,y)$ through
the center of the dot. We  use two approximate models, the simplest
one being an infinite square well, of width $w_y \approx  350 \ nm$.
The slightly fancier one is a truncated harmonic oscillator: 
 \begin{eqnarray}
W_{tho}(y) &=&  \qquad 0 \hskip 2.8cm {\rm (flat} \,\,\, {\rm bottom)}
\nonumber \\
  &=& -0.13 + {1\over 2} k_y (y - y_0)^2   \quad {\rm (walls)} \ .
\label{eq:14}
\end{eqnarray}
with $ y_0 = 238 \ nm$ and $k_y = 7.35 \cdot 10^{-6} \ nm^{-2} $.
As can be seen in Fig. \ref{fig5} this  parametrization (plus the
constant term $U_0$) reproduces the main features of the $x=0$
sections of the PTF potentials. 

By combining eqs. \ref{eq:6} to \ref{eq:14}  we determine a separable 
analytic potential model for the dot containing a desired number $Q$ 
of electrons. This removes the necessity of repeatedly solving the 
PTF equations for the self-consistent field, while studying the decay 
process. 

\subsection{Quasibound states of the dot}
We construct the electron wave functions inside the dot in the 
envelope function approximation, using our parametrized potential, 
$U_s(x,y)$. The single electron energies are 
 \begin{equation}
E_{n_x,n_y} = E_{n_x} + E_{n_y} \,
\label{eq:15}
\end{equation}
and the electron wavefunctions factorize
 \begin{equation}
\Psi_{n_x,n_y}(x,y) = \phi_{n_x}(x) \psi_{n_y}(y) \, .
\label{eq:16}
\end{equation}
The factors satisfy 1D  Schr\"odinger equations:
 \begin{eqnarray}
&-& {\hbar^2\over {2m^*}} \phi_{n_x}''(x) + U(x) \phi_{n_x}(x) =
E_{n_x} \phi_{n_x}(x) \nonumber \\
&-& {\hbar^2\over {2m^*}} \psi_{n_y}''(y) + W(y) \psi_{n_y}(y) =
E_{n_y} \psi_{n_y}(y)
 \label{eq:17}
\end{eqnarray}
The second equation is for a confined wavefunction, easily
solved  by standard numerical methods. We label the solutions by
the number of loops, $n_y$, of the eigenfunction. For example, taking
$W(y)$ to have hard walls, the energy is
 \begin{equation}
E_{n_y,sw } = {\hbar^2\over {2m^*}} \left({{n_y \pi}\over
w_y}\right)^2 \ .
\label{eq:18}
\end{equation}
For the truncated harmonic oscillator shape there is no similar 
analytic expression, but the dependence on $n_y$ is similar.

    The $x$-dependent equation describes 1D electrons confined in the
dot by the ``leaky barriers''. Weakly quasibound state solutions
were computed using methods described in \cite{K77}. However, for levels
corresponding to the long tunnelling lifetimes observed in the
experiment, the energies and eigenfunctions can be computed well
enough by the simpler prescription of setting the electron
wavefunction to zero at the points $\pm x_b$ inside the barriers.
Furthermore, if only the eigenvalues and lifetimes are needed, we have 
checked that the WKB quantization condition is adequate:
 \begin{equation}
\int_{x_l}^{x_r} \sqrt{{{2m^*}\over \hbar^2} ( E(n_x) - U(x) )} \  dx =
\left(n_x - {1\over 2} \right) \pi
\label{eq:19}
\end{equation}
In the Appendix we describe the determination of the lifetimes, 
$\tau_{n_x}$. From here on the energies presented are obtained in the 
WKB approximation. The differences from the more accurate predictions 
using the true quasibound state energies can scarcely be seen on the 
scale of the graphs. For barrier penetrability we use eq. 
\ref{eq:12}. 

We ``construct" the desired dot 
configuration with excess electrons by generating a $U_s(x,y)$ for 
the chosen value of $Q$, and  filling the levels as follows: 
a) First we list the $(E_{n_x},\tau_{n_x})$, in 
order of  increasing $n_x$ (and therefore of increasing energy and
decreasing lifetime.) This list is truncated at an $n_x = n_{x,max}$
whose lifetime is less than $0.01 \ sec.$
b) Next we form a list of 2D levels $(n_x,n_y)$  by choosing those for which
 \begin{equation}
E_{n_x} + E_{n_y} \le E_{n_{x,max}} + E_{n_y = 1} \ .
\label{eq:20}
\end{equation}
The levels in this list are occupied in order of increasing energy 
and according to Fermi statistics, see eqs. \ref{eq:a5} \ref{eq:a6}. 
We choose the dot Fermi level so that the number of electrons is the 
desired Q. It is supposed that, for the long lifetimes observed in 
the experiment, the electrons have time to lose energy by phonon 
collisions and occupy the quasibound states of lowest energy. Then, 
as described in Appendix A, we determine the lifetime for one electron 
to escape from the dot.  This involves a weighted average of 
the level lifetimes, according to the occupancy of each level at the 
experimental temperature, $T' = 100 \ mK$. 

To produce a sequence of decays for comparison to experiment we 
proceed as follows: {\it i)} we start with a dot containing  a number 
of electrons, $Q_0$, chosen large  enough so that the  lifetime for 
one electron to escape is smaller than those observed in experiment. 
{\it ii)} We redetermine the barrier and dot configuration for $Q = 
Q_0 -1$ electrons, as described in the above paragraph and determine 
again the corresponding lifetime for escape of one electron. This 
process is repeated to generate a sequence of decays that covers and 
extends beyond the range of lifetimes measured in experiment. From 
that list we choose as the first observed electron decay that 
corresponding to the $Q$ whose lifetime is the first to be larger 
than $t_0 = 25 $ seconds. 

\begin{figure}
\leavevmode
\epsfxsize=8cm
\epsffile{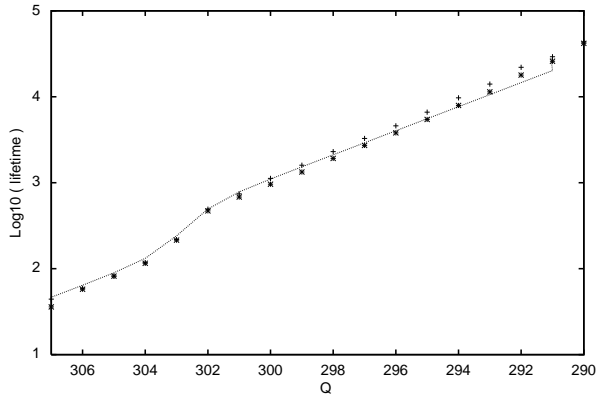}
\caption{  Calculated lifetimes (in seconds) 
when $W(y)$ is either the truncated harmonic oscillator: stars, or a 
square well with $w_y = 380 \ nm$: $+$ signs . The dotted line is the 
prediction of the two level model, eq. \protect{\ref{eq:tm2}}. 
 } 
\label{fig6}
\end{figure}

\section{Results and discussion}

In Fig. \ref{fig6} we show results from our model, using 
parameters chosen as described above, for a range of lifetimes 
extending over three orders of magnitude. The stars correspond to the 
truncated harmonic oscillator choice for $W(y)$, whereas the $+$'s 
are for the square well choice (with a value $w_y = 380 \ nm$ 
chosen to optimize the agreement with the other prescription in the 
range of experimental lifetimes, from 10 to 1000 seconds.) One 
sees that the trends are very similar. For $Q$ in the 
neighbourhood of $304$, the predicted decay lifetimes fall in the 
experimental range. 

As already mentioned in Section II, our PTF 
simulations predict $Q=286 $ for the dot in equilibrium with the 
surrounding electron gas. This is also what we find with this 
separable model, as the curve of lifetimes shown in figure 
\ref{fig6} extrapolates smoothly up to $Q= 287$, for which we predict 
a lifetime of $\log_{10}\tau = 5.2$, or 44 hours. After that, the Fermi 
level of the electrons inside the dot falls below that of the 
surrounding 2DEG and further decays are blocked. It should take 
almost two days for the dot to reach equilibrium with its surroundings.  

Before attempting a more detailed comparison with the 
experimental data it is useful to examine the main features in our 
predicted sequences. First we focus on the linear behaviour 
for values $ Q < 300$. (We have found similar behaviour in 
other ranges of $Q$ when we use slightly different sets of 
parameters.) Such linear dependence occurs when our model produces a 
sequence of decays dominated by those from a single 1D electron 
level; {\it i.e.} corresponding to a fixed value of $n_x$. To understand 
why, suppose that at zero temperature and for $Q$ electrons, the 
occupied level with shortest lifetime is $(n_{x,s},n_y)$, and that 
$\{n_x',n_y'\}$ are occupied levels with higher energy and longer 
lifetime (this requires that at least $n_x' < n_{x,s}$ for longer 
lifetime and $ n_y' > n_y$ for higher total energy.) When one forms 
the $Q-1$ electron configuration according to the rules explained 
above, one of the $\{n_x',n_y'\}$ levels will be empty, whereas the 
level $(n_{x,s},n_y)$ will again be filled. In more physical terms: 
all the electrons with energy above that of the level with shortest 
lifetime will lose energy by phonon collisions and fall into 
the leaky level, from which they finally escape. Since the 
lifetime does not depend on $n_y$, all the electrons with energy 
above that of the state $(n_{x,s}, n_y = 1)$ will escape through the 
same leaky 1D level, $n_{x,s}$, which remains the favoured decay 
channel as long as it is occupied. 
Therefore the total probability for {\bf one} electron to escape from 
the occupied states  with quantum number $n_{x,s}$ is the probability 
for a single 1D  electron with energy $E_{n_{x,s}}$, multiplied by  
the number of electrons in occupied states with the same quantum 
number $n_{x,s}$: $q_{n_{x,s}}$: 
 \begin{equation}
\tau(Q) = {{\displaystyle{\tau_{n_{x,s}}(Q)}} \over 
               {\displaystyle{q_{n_{x,s}}(Q)}}} \quad ,
\label{eq:tm1}
\end{equation}
and when the occupation $q_{n_{x,s}}$ of the leaky level is 
constant, the linear variation of $\log_{10}(\tau)$ reflects that of 
the lifetime of the leaky level. This is where the 2D nature of the 
quantum dot asserts its presence, even though the decay appears to 
proceed only in one dimension.

\begin{figure}
\leavevmode
\epsfxsize=8cm
\epsffile{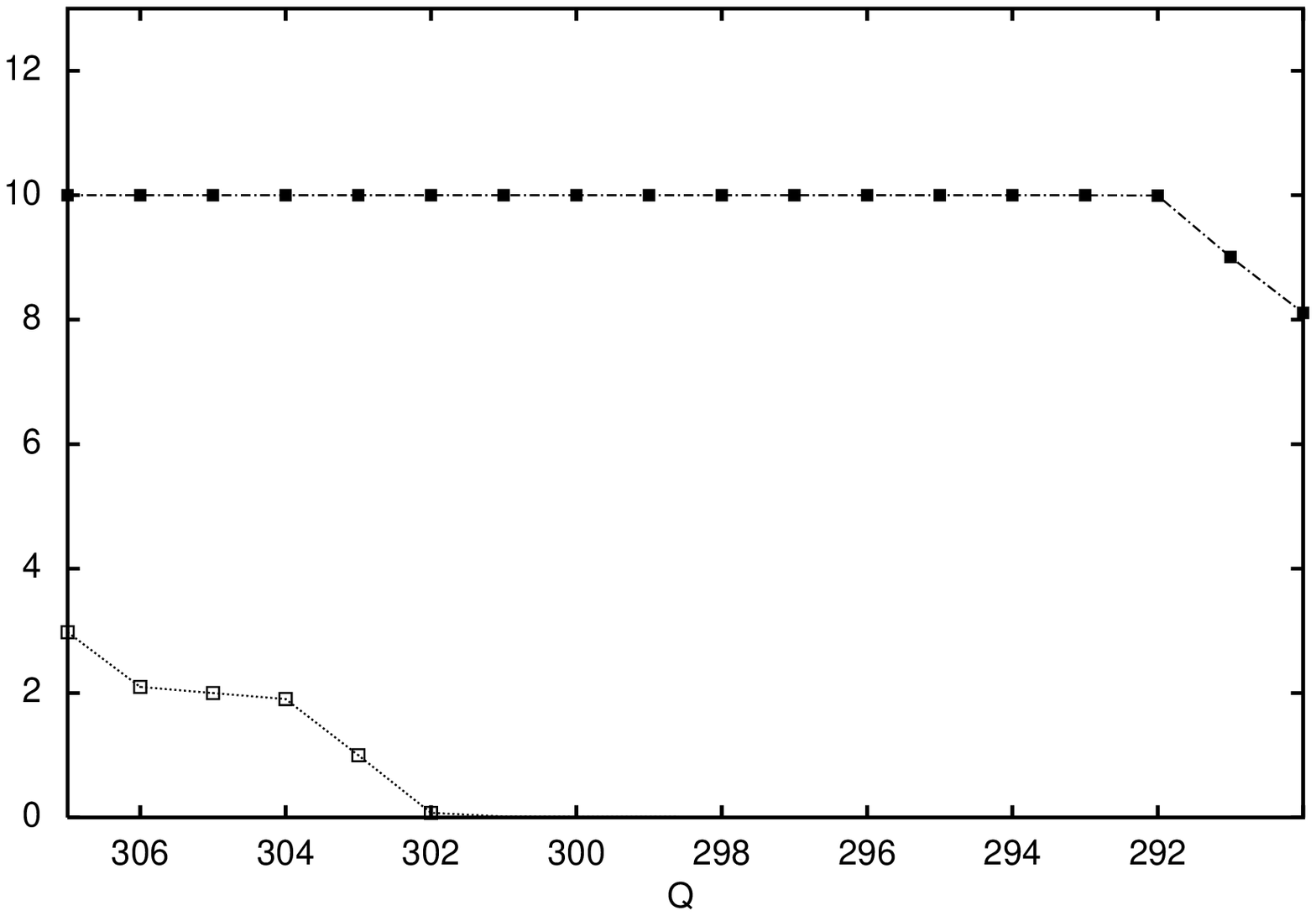}
\caption{Total occupation of levels with $n_x = 13$ (black squares) 
and $n_x = 14$ (open squares). The lines are drawn to guide the 
eye. The truncated harmonic oscillator model was used for $W(y)$. 
 } 
\label{fig7}
\end{figure}

In Fig. \ref{fig7} we show the occupations of the two levels with the 
shortest lifetimes. One sees that when $Q < 300$ the occupation of 
the $n_x = 13 $ level stays practically constant and $n_x = 14 $ 
level remains empty. For higher values of $Q$ both levels contribute 
significantly to the escape lifetime. In this situation: 
 \begin{equation} 
\tau(Q) = {1\over{\displaystyle{ {{q_a(Q)}\over{ \tau_a(Q)}} + {{q_b(Q)}
\over {\tau_b(Q)}}}}}\ .
\label{eq:tm2}
\end{equation}
This is shown as the dotted curve in Fig. \ref{fig6}, and it accounts 
very well for the trend of the lifetimes predicted by the separable 
model. 

\begin{figure}
\leavevmode
\epsfxsize=8cm
\epsffile{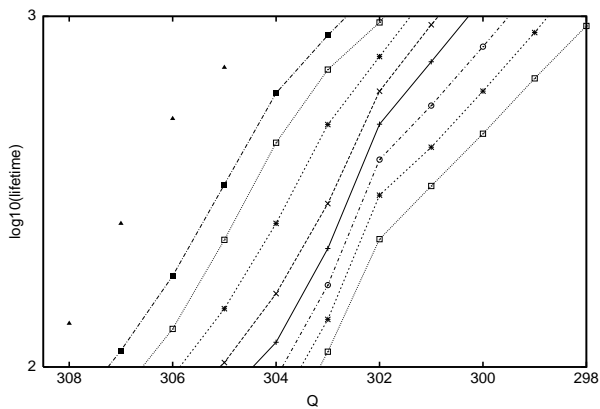}
\caption{ Slow decays: Calculated lifetime sequences corresponding to 
variations of  the standard barrier width from $-3 \%$ (right) to 
$+4\%$ (left) in steps of $1 \%$ . The $+$ signs joined by a 
continuous line (to guide the eye) correspond to the prediction for 
the standard set of parameters. Experimental points (left triangles) 
taken from Fig. 1 with the origin of $Q$ shifted arbitrarily. } 
 \label{fig8} 
\end{figure}

Our separable model favours the appearance of the linear decay 
sequences, because of the degeneracy in lifetime of states with the 
same $n_{x,s}$. A non-separable model would lift that degeneracy and 
then the lifetime sequences should show a behaviour intermediate 
between the two situations discussed above. In particular, the sudden 
change of slope at  $Q=302$ in Fig. \ref{fig6} would presumably 
spread over a wider range of values of $Q$. 
Not surprisingly, the predicted lifetimes for the observed decays 
depend sensitively on details of the barrier shape. Those shown in 
Fig. \ref{fig8} correspond to the square well choice for $W(y)$, and 
our standard set of parameters. In addition we show how the lifetimes 
vary when the barrier width  is changed by amounts ranging from $+ 
4\%$ to $-3\%$ (from left to right). As can be seen, the exact value 
of each decay lifetime depends quite strongly on the barrier width, 
as expected for a tunnelling process. But the number of slow decays 
is much more stable: four or five in most of the cases shown, and in 
several cases their lifetimes are quite compatible with the 
experimental points. In particular it is remarkable that a $2\%$ 
increase in the standard barrier width produces a sequence (third 
line from left) in excellent agreement with experiment (disconnected 
points shifted to extreme left). 

There is a clear distinction between the lifetime trends of the 
thicker and thinner barrier widths. In the latter one sees very 
clearly the transition between escape from the $n_x = 13 $ and the 
$n_x = 14 $ levels at $Q = 302$. For the thicker barrier widths, 
escape is dominated by the $n_x = 14$ levels that become 
progressively more occupied above $Q = 302$.

\begin{figure}
\leavevmode
\epsfxsize=8cm
\epsffile{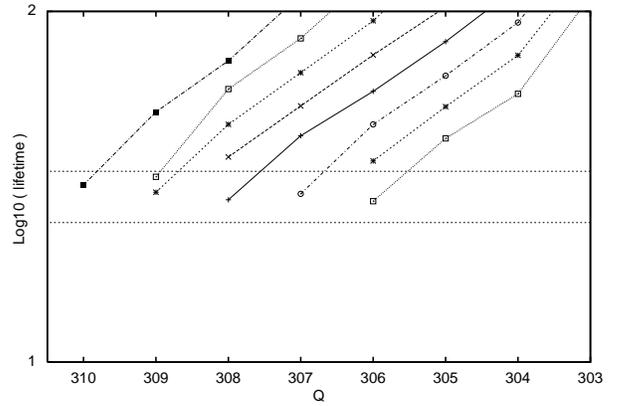}
\caption{ Same as Fig. 8, but for the fast decays.  } 
 \label{fig9} 
\end{figure} 

We have explored the dependence of the model predictions on changes 
by similar percentages of the barrier heights, potential floor $U_0$, 
and the width, $ w_y$, of $W(y)$. The results are qualitatively 
similar to those shown above for the changes in the barrier width, 
with a number of slow decays ranging from 4 to  6, and in some 
cases they are very similar to the data in 
Fig. \ref{fig1}. We therefore conclude that our model predictions are 
quite consistent with the experimental trends, although a 
quantitative comparison with the measured  lifetimes is hampered by 
the strong sensitivity of tunnelling to any small change in the 
barrier shape. 

{}Finally we show in Fig. \ref{fig9} predictions for the fast decays: 
their number and location in a graph such as that of figure 
\ref{fig1} depends very sensitively on the time ($t_0$) beyond which 
the experiment measures lifetimes. That is, as the dot is isolated 
there must be a burst of very short lived escapes, but after some 
seconds one reaches the stage where separate events can be recorded 
and the lifetimes deduced. The two dashed horizontal lines in Fig. 
\ref{fig9}  correspond to values of $\log_{10}( 25 \ s)$ and 
$\log_{10}(35 \ s)$. As can be seen, when we explore the same range 
of barrier widths as in Fig. \ref{fig8}, the number of fast decays 
above that $t_0$ varies from 3 to 4. The overall trend seems to be 
consistent with experiment, in particular if the value of $t_0$ is 
increased towards the more pessimistic estimate of $35 \ s$.  

\section{Summary and Conclusions}

Electron escape from a strongly isolated dot with excess electrons 
has been studied in the framework of the self-consistent 
Poisson-Schr\"odinger and Poisson-Thomas-Fermi approximations. Based 
on these calculations a rectangular separable potential model has 
been devised which incorporates the main features of the 
self-consistent field. Rearrangement effects are taken into account 
by recalculating the confining potential $U_s(x,y)$ after each 
electron escape. 

The use of a separable potential introduces certain correlations in 
the energy spectrum of the single electron orbitals. A more realistic 
confining potential would have a more rounded shape, which would 
remove the separability, and modify those correlations.  In 
the same vein, the tunnelling in our simplified model is 1D, whereas 
the actual process is 2D. 

We find it quite remarkable that despite all these simplifications the 
predictions turn out to be so satisfactory. The model therefore may 
be reliable for extrapolating to longer times. For instance we find 
that the isolated dot would hold one excess electron for as long as 44 
hours. On such a time scale, one could use well isolated dots 
containing a few long lived electrons, to study their entangled 
states. This would open an interesting new approach to the 
implementation of quantum computation in semiconductor devices.

\acknowledgements

We are grateful to DGES-Spain for continued support through grants 
PB97-0915 and UE97-0014(JM), to EPSRC-UK (CS) and to NSERC-Canada for 
research grant SAPIN-3198 (DWLS). This work was carried out as part 
of QUADRANT, Esprit project EP-23362 funded by the EU.

\appendix
\section{ Lifetimes}
We summarize here the expressions relating the lifetimes to the 
probability of transmission across the barrier. We follow the 
standard treatment and definitions for alpha decay in nuclear 
physics, as can be found for example in ref.\cite{PB75}. 

Our potential $U_s(x,y)$ is separable, and the electron can escape only
across the barriers in the $x$-direction. Therefore, we have adapted
the expressions derived in \cite{PB75} to the 1D situation.

  The lifetime $\tau = 1/\lambda$ is the inverse of the 
``decay constant'', defined as the number of ``decays'' per second 
per parent ``dot''. For one dot the electron wavefunction is 
normalized to unity over the volume inside the barriers, and 
$\lambda$ for a given level is just the outgoing flux at large 
distance. 

When the decay probability is small, one can treat the 
electron as confined in the dot. Classically, its trajectory will
oscillate between the right, $x_r$, and left, $x_l$  turning points,
with a period
 \begin{equation}
P = 2 \int_{x_r}^{x_l} {{ dx}\over v(x) } \ ,
\label{eq:a1}
\end{equation}
where $v(x)$ is the classical electron velocity at energy $E_x$:
 \begin{equation}
v(x) = \sqrt{ {2\over m^*}\left( E_x - U(x) \right)} \, .
\label{eq:a2}
\end{equation}
The flux $\lambda$ is then given by the frequency of hits against the 
barriers, $2/P$, times the transmission probability $T$ across a 
barrier, and therefore:
 \begin{equation}
\tau = {1\over \lambda} =  \frac{1}{T} \int_{x_l}^{x_r}
{{dx} \over v} \ .
\label{eq:a3}
\end{equation}
This expression is very convenient because the transmission
coefficient eq. \ref{eq:12} for our parametrized potential, $U(x)$, is known
analytically \cite{MF53}. For more general barrier profiles and the
long lifetimes of interest, one can use the WKB approach and its
corresponding connection formulae across the barrier (see e.g.
Appendix D of \cite{PB75}):
 \begin{eqnarray}
T_{WKB} &\approx& e^{ 2 \omega} \nonumber \\
\omega &=& \int_{x_r}^{x_t} \kappa \ dx = \int_{x_r}^{x_t} \sqrt{
{{2m }\over \hbar^2} \left( U(x) - E_x \right) }\, \, dx
\label{eq:a4}
\end{eqnarray}

If the WKB wave function is used inside the well to determine the
period $P$, the same decay half-life is obtained as in eq.
\ref{eq:a3} above.

Since the dot is located inside a crystal at temperature $T'$, via
phonon coupling the electrons in the dot should also be at the
same temperature. The level occupations $f(E)$ are determined by 
Fermi statistics: 
 \begin{equation}
f(E) =  \bigl[ {1 + e^{{ E-E_F} \over {k_BT'}}} \bigr]^{-1} \ ,
\label{eq:a5}
\end{equation}
where these are now 2D energies. The Fermi level is obtained from 
 \begin{equation}
Q = \sum_{i=(n_x,n_y)} \ 2 f(E_i)  ,
\label{eq:a6}
\end{equation}
where the $2$ accounts for spin degeneracy. For the ensemble
of electrons in the dot, the flux $\lambda$ will now be the
sum of fluxes for each occupied single particle level, weighted by
the level occupancy:
 \begin{equation}
\lambda = \sum_{i=(n_x,n_y)}  2 f(E_i) \lambda_i
\label{eq:a7}
\end{equation}
and the corresponding half-life is still $\tau = 1/\lambda$. 
In particular this argument applies in the $T'=0$ limit, as we 
implicitly assumed in  Section II to explain the sequence of 
lifetimes.

\section{Lifetime dependence on $Q$ }

For a level of given $n_x$, the lifetime depends on $Q$ because the 
barrier characteristics change, and so does the level energy, 
$E_{n_x}$. The latter varies mainly because $U_0$ depends on Q, and 
this affects the transmission probability $T$. To good approximation 
 \begin{equation}
{{ d E_{n_x}}\over {d Q} } \simeq {{d U_0}\over {d Q}} 
\label{eq:b1}
\end{equation}

Neglecting the dependence of the level lifetime on the period $P$, we 
can write 
 \begin{eqnarray}
\frac{ d \,\ln \tau }{dQ}
&\simeq& - {{ d \ln   T}\over { d E}} {{dE} \over {dQ}}  \ .
 \label{eq:b2}
\end{eqnarray}
Taking the transmission probability $T$ from the WKB expression leads
to
 \begin{eqnarray}
{{ d \ln \tau }\over {dQ}} &=& 2 {d \over {dQ}}
\int_{barrier}
\sqrt{{{2m^*}\over \hbar^2} (U(x) -E) }\ dx \
 \nonumber \\
&=&     \int_{barrier} \sqrt{{{2m^*}\over \hbar^2}
{1\over {U(x) -E}}} \left( {{dU(x)}\over{dQ}}\, -\, {{dE}\over{dQ}} \right)
dx  \nonumber \\
\label{eq:b3}
\end{eqnarray}

Noting eq. \ref{eq:b1}, the second contribution to the integral
depends linearly on the placement of the potential floor. However, the 
variation of the barrier shape ($dU(x) / dQ$) cannot be neglected. 
Indeed, eq. \ref{eq:8} gives approximately 
 \begin{eqnarray}
{{dU(x)}\over {dQ}} &=& {dU_0\over dQ} \left( {1\over 3} + {2\over 3}
{{\sinh^2 \left({{x-x_b}\over w_b}\right)}\over {e^{-
2\mu}\cosh^2\left({{x-x_b}\over w_b} - \mu\right)}}\right) 
\nonumber \\ && \hskip 2cm 
{\rm when} \quad x < x_b \nonumber \\
{{dU(x)}\over {dQ}} &=& {dU_0\over dQ} \left( {1\over 3} - {1\over 3}
{{\sinh^2 \left({{x-x_b}\over w_b}\right)}\over \phantom{-}
{e^{2\mu}\cosh^2\left({{x-x_b}\over w_b} - \mu\right)}} \right) 
\nonumber \\ && \hskip 2cm 
{\rm when} \quad x > x_b \ .
 \label{eq:b4}
\end{eqnarray}
Using \ref{eq:b4}, the contribution from $dU/dQ$ to the integral of 
eq. \ref{eq:b3} is obtained with an accuracy better than $ 2\%$. 

\begin{figure}
\leavevmode
\epsfxsize=8cm
\epsffile{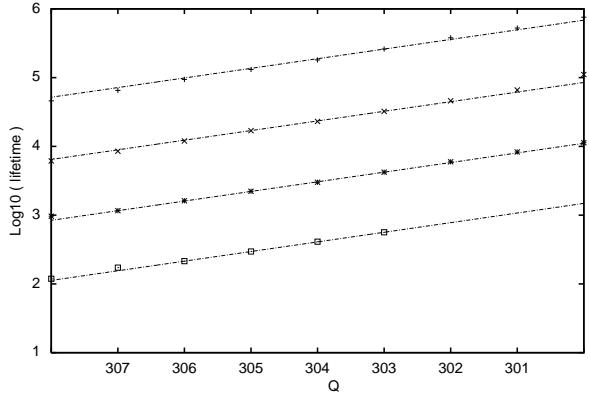}
\caption{ Dependence of lifetime on $Q$ for the occupied levels with 
$n_x = 11$: $+$ signs; $n_x = 12$: $\times$ signs; $n_x = 13$: stars 
and $n_x = 14$ open squares. The dash dotted lines correspond to eq. 
\protect{\ref{eq:b5}}. } 
 \label{fig10} 
\end{figure} 

\noindent
{}For the standard choice of parameters, and $Q$ in the range $300$ to $310$
the computed values of $d \log_{10} \tau / dQ$ turn out to be $\simeq -0.14$
for the levels of interest.  In Fig. \ref{fig10} we plot the evolution of the
level lifetimes with $Q$, compared to the expression ($Q_0 = 303 $)
 \begin{equation}
\log_{10} \tau_{n_x}(Q) = \log_{10} \tau_{n_x}(Q_0) -0.14 ( Q - Q_0 ) 
\ .
\label{eq:b5}
\end{equation}

\end{document}